\documentclass[twocolumn]{aastex62}
\usepackage{amsmath,amstext}
\usepackage{savesym}
\savesymbol{tablenum}
\usepackage{apjfonts} 
\usepackage{verbatim}
\usepackage{chngcntr}
\usepackage{siunitx}
\newcommand{\be}{\begin{eqnarray}}
\newcommand{\ee}{\end{eqnarray}}

\begin{document}

\normalsize


\title{\Large \textbf{Probing Planets with Exomoons: The Cases of Kepler-1708 b and Kepler-1625 b}}

\author{Armen Tokadjian}
\affiliation{Department of Physics and Astronomy, University of Southern California, Los Angeles, CA 90089-1342, USA; tokadjia@usc.edu}
\affiliation{The Observatories of the Carnegie Institution for Science, 813 Santa Barbara St., Pasadena, CA 91101, USA}

\author{Anthony L. Piro}
\affiliation{The Observatories of the Carnegie Institution for Science, 813 Santa Barbara St., Pasadena, CA 91101, USA}

\begin{abstract}
The tidal interactions between a planet and moon can provide insight into the properties of the host planet. The recent exomoon candidates Kepler-1708 b-i and Kepler-1625 b-i are Neptune-sized satellites orbiting Jupiter-like planets and provide an opportunity to apply such methods. We show that if the tidal migration time is roughly equal to the age of these systems, then the tidal dissipation factor $Q$ for the planets Kepler-1708 b and Kepler-1625 b have values of $\sim3\times10^5-3\times10^6$ and $\sim1.5\times10^5-4\times10^5$, respectively. In each case, these are consistent with estimates for gas giant planets. Even though some work suggests an especially large semimajor axis for Kepler-1625~b-i, we find that this would imply a surprisingly low $Q\sim2000$ for a gas giant unless the moon formed at essentially its current position. More detailed predictions for the moons' initial semimajor axis could provide even better constraints on $Q$, and we discuss the formation scenarios for a moon in this context. Similar arguments can be used as more exomoons are discovered in the future to constrain exoplanet interior properties. This could be especially useful for exoplanets near the sub-Neptune/super-Earth radius gap where the planet structure is uncertain.


\end{abstract}


\section{Introduction}


The study of moons is crucial to our understanding of extrasolar planetary systems. They provide information about the formation and evolution of such systems \citep{morbidelli16}, they impact the climate and habitability of planets, and may even be sites for life themselves \citep{heller14}. The large number of moons in our own Solar System suggests that many exomoons also reside around exoplanets. But despite efforts to detect these exomoons (e.g., \citealp{brown01}; \citealp{kipping12}), their discovery has been challenging. 

Nevertheless, recently some exomoon candidates are beginning to be found. The first was Kepler-1625~b-i \citep{teachey}, which was reported to show timing variations and transit signatures consistent with a large Neptune-sized moon (although for further discussions about the viability of this exomoon, see \citealp{kreidberg19}; \citealp{teachey20}). More recently, there has been the discovery of Kepler-1708~b-i \citep{kipping22}. This system consists of a Jupiter-sized planet orbiting a Sun-like star, which showed a moon-like transit signature consistent with a $\sim2.6\,R_{\oplus}$ satellite. 

Just as the tides between the Earth and Moon slow the Earth's spin and cause the Moon to migrate outward, tides in these systems will impact their architecture. This has inspired a number of recent theoretical studies of this dynamics (e.g., \citealp{barnes2002}; \citealp{sasaki12}; \citealp{sasaki14}; \citealp{adams16}; \citealp{Piro18}; \citealp{Tokadjian}; \citealp{quarles20}). Inversely, now that there is a growing number of candidate exomoons, their observed architectures can be used to constrain the strength of the tides in these systems. This is closely related to the properties of the moon-hosting planet, providing important information about its inner structure.


Motivated by these possibilities, we study the tidal interaction between Kepler-1708 b and Kepler-1625 b and their potential exomoons. By invoking both analytic and numerical methods, we explore the range of tidal dissipation rates plausible for the planet given the current parameters described for the exomoons. In this way, we can constrain the interior properties of the planet. By analyzing the history of the moon's migration, we formulate possible initial separations between planet and moon, which gives clues to the formation pathway that the moon followed in its past.
 
In Section~\ref{sec:analytic}, we introduce a simple tidal lag model and apply it to the Kepler-1708 and Kepler-1625 systems to analytically estimate tidal dissipation parameters of the host planet. Then in Section~\ref{sec:numerical}, we numerically analyze the range of these parameters we can expect given different initial scenarios for planet-moon configuration. We also discuss possible formation mechanisms for the moon. We provide a summary and conclusion in Section~\ref{sec:conclusion}.

\section{Tidal Dynamics Equations}
\label{sec:analytic}

We start by presenting the formulism we use to address the tidal dissipation between a single planet, moon, and star. The star and the moon induce a tidal bulge on the planet which is out of phase with the lines joining the centers of the bodies due to tidal lag. The resulting torque will slow the spin of the planet and push the moon away (e.g., \citealp{counselman}). Since we consider the secular evolution of the system, we assume that tides from the moon and star act independently on the planet once averaged over long timescales. \citep{Piro18}. We use a parameterized model with constant phase lag (CPL) to calculate migration rates of the exomoon candidate.

\subsection{CPL Model}
\label{cpl}
Following the basic prescription in \citet[and references therein]{Tokadjian}, the time evolution of single planet, moon, and star can be written as a set of differential equations which describe the spin, $\Omega_p$ of the planet,
\be
	\frac{d}{dt}(I_p\Omega_p)=N_s + N_m,
	\label{spin}
\ee
the change in the planet's orbital separation $a_s$,  
\be
	\frac{d}{dt}[M_p(GM_sa_s)^{1/2}]=-N_s,
	\label{seps}
\ee
and the change in the moon's orbital separation $a_m$,
\be
	\frac{d}{dt}[M_m(GM_pa_m)^{1/2}+I_m n_m]=-N_m.
	\label{sepm}
\ee
The masses of the star, planet, and moon are $M_s$, $M_p$, and $M_m$, respectively, and the moments of inertia for planet and moon are given by $I_p$ and $I_m$, respectively. The orbital frequency of the moon is $n_m$, and $N_s$ and $N_m$ represent the torque on the planet due to the star and moon, respectively. The second term on the left hand side of Equation~(\ref{sepm}) is due to tidal locking of the moon to the planet.


In the CPL model, the torques can be written as \citep{efroimsky13}

\be
	N_s = \frac{3}{2}\frac{k_2}{Q}\sigma_s\frac{GM_s^2R_p^5}{a_s^6},
	\label{CPLtorques}
\ee
and
\be
	N_m = \frac{3}{2}\frac{k_2}{Q}\sigma_m\frac{GM_m^2R_p^5}{a_m^6},
	\label{CPLtorquem}
\ee
where $Q$ is the quality factor $Q$ that describes the interior structure of the planet (e.g., \citealp{efroimsky}) and $k_2$ is the Love number describing the rigidity of the planet and is taken to be 0.3 for rocky planets like the Earth \citep{Yoder1995} and 0.38 for gas giants like Jupiter \citep{Gavrilov}. These equations have the relation $N\propto R^5$, which corrects the typo in Equations~(8) and (9) of \citet{Tokadjian} that have $N\propto R^2$. Note that $\sigma_s = \mathrm{sgn}(n_s-\Omega_p)$ where $n_s$ is the orbital frequency of the planet around the star and $\sigma_m = \mathrm{sgn}(n_m-\Omega_p)$. In this study, $\sigma_s=\sigma_m=-1$ since we assume the planet is spinning at a faster rate than it is orbiting the star and being orbited by the moon.

\subsection{Estimating Q}
Before delving into numerical integration of the differential equations in the previous subsection, it is helpful to analytically estimate how the quality factor $Q$ depends on the parameters of the system. We do this by inserting Equation~(\ref{CPLtorquem}) into Equation~(\ref{sepm}) to derive a differential equation for $da_m/dt$. This is then integrated from $a_{m,0}$ at $t=0$ to $a_m$ at $t=\tau_{\mathrm{mig}}$, the total time the moon has been migrating. We obtain,

\be
    \tau_{\mathrm{mig}} = \frac{2QM_p^{1/2}}{3k_2G^{1/2}R_p^5M_m}\left[\frac{a_m^{13/2}-a_{m,0}^{13/2}}{13} - \frac{2(a_m^{9/2}-a_{m,0}^{9/2})}{15}R_m^2\right],\;\;\;
\ee
where the second term scales as $a_m$ to the 9/2 power due to the tidal locking of the moon as mentioned above. Taking the limit that $a_m\gg a_{m,0}$ (that the current position of the moon is far from where it started) and $a_m^2\gg R_m^2$, we simplify the expression for the migration timescale to be

\be
    \tau_{\mathrm{mig}} = \frac{2QM_p^{1/2}a_1^{13/2}}{39k_2G^{1/2}R_p^5M_m},
    \label{tmigf}
\ee
consistent with the form derived in \citet{quarles20}.

By solving Equation~(\ref{tmigf}) for $Q$, we obtain a simple expression for the quality factor of a planet given the migration time (or age) of the moon and its current position,

\be
\begin{aligned}
    Q = 3.8\times10^5 \left(\frac{k_2}{0.38}\right)\left(\frac{\tau_{\mathrm{mig}}}{\mathrm{5Gyr}}\right)\left(\frac{M_m}{5M_{\oplus}}\right)\left(\frac{R_p}{R_{J}}\right)^5\\ \times\left(\frac{M_p}{M_{J}}\right)^{-1/2}\left(\frac{a_m}{10^{11} \mathrm{cm}}\right)^{-13/2}.
    \label{QQ}
\end{aligned}
\ee
Checking the values for the Earth-Moon system with $\tau_{\mathrm{mig}}$ set to 4.5\,Gyr, the approximate age of the moon, we obtain $Q\approx30$ which is reasonably consistent with the expectation for a rocky planet. We also apply this equation to the Jupiter-Io system and calculate $Q\approx3 \times10^5$ for Jupiter, an acceptable value for gas giant planets.


To apply these equations to exomoon candidate Kepler-1708 b-i, we use $M_s=1.1\,M_{\odot}$, $R_s=1.1\,R_{\odot}$, $a_p=1.64$\,AU, $R_p=0.89\,R_J$, $a_m=11.7\,R_p=5.54\,a_\mathrm{R}$, and $R_m=2.6\,R_{\oplus}$ \citep{kipping22}, where 

\be
    a_\mathrm{R} = 2.16R_m\left(\frac{M_p}{M_m}\right)^{1/3},
    \label{roche}
\ee
is the Roche-lobe radius \citep{frank}. Because only upper bounds of the planet and moon mass are provided, we assume $M_p=0.81\,M_J$, which roughly matches the density of Jupiter, and $M_m=5\,M_\oplus$, which places the moon in the sub-Neptune regime with approximately the density of Neptune. The age of the system is estimated to be about 3.16\,Gyr. Plugging these values into Equation~(\ref{QQ}), we obtain a $Q$ of just over $10^6$ which places the planet Kepler-1708~b in the expected category of a gas giant similar in structure and composition to Jupiter.

We similarly apply Equation~(\ref{QQ}) to the Kepler-1625 system. The parameters are $M_s=1.1\,M_{\odot}$, $R_s=1.8\,R_{\odot}$, $a_p=0.88$\,AU, $R_p=1.04\,R_J$ and $R_m=4.9\,R_{\oplus}$, with an age of $8.7\pm2.1$\,Gyr. The current semi-major axis of the moon is not well constrained, so we adopt the approximation provided by \citet{teachey18} and take $a_m=19.1\,R_p=5.7\,a_\mathrm{R}$. For the masses, we assume $M_p=3\,M_J$ which is the most probable value suggested by photodynamical modeling \citep{teachey20}, and $M_m=19\,M_\oplus$, so that the exomoon is Neptune-sized.  We calculate a $Q$ of about $2\times10^5$, similar to the Jupiter-Io system. Although the planet mass is poorly constrained, this will not affect the results significantly due to the weak dependence of $Q$ on $M_p$ in Equation~(\ref{QQ}).


\section{Detailed Analysis and Discussion}
\label{sec:numerical}

In the previous section, we derived rough values for $Q$ given the best estimate of the age of the systems. But given that there is uncertainty in both the age of the planet as well as the age of the moon, here we consider the implied $Q$ for a range of reasonable ages. We then discuss possible moon formation pathways and the likely $Q$ of the planet given the nature of each scenario.

\subsection{Kepler-1708}

\begin{figure}
\epsscale{1.0}
\plotone{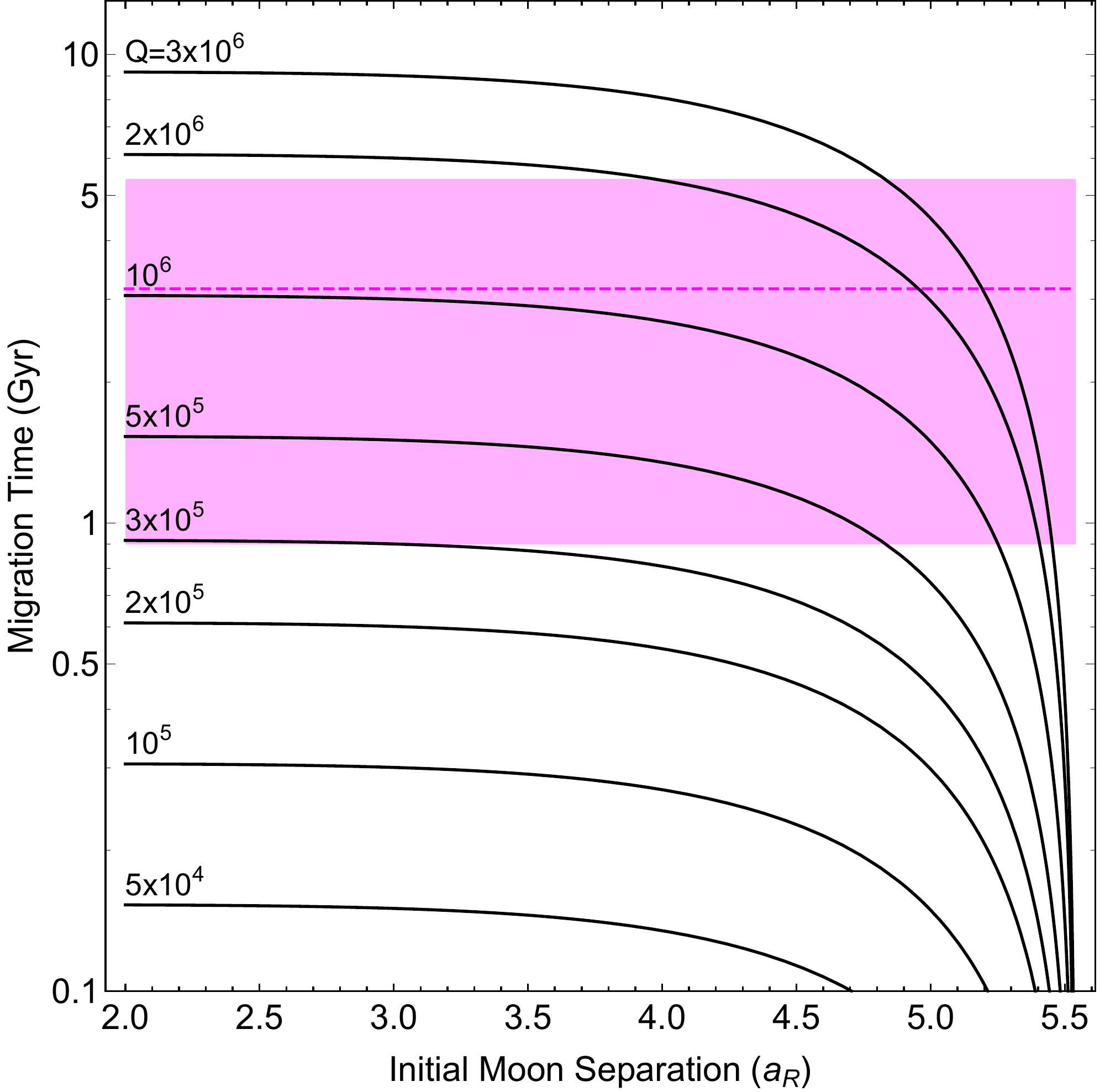}
\caption{Timescales for exomoon candidate Kepler-1708 b-i to migrate to its current estimated separation of $a_m=5.5\,a_\mathrm{R}$. Each black curve corresponds to a constant $Q$ in the CPL model and is shown above the respective curve. The magenta dashed line is the best estimate for the age of the system, 3.16\,Gyr, while the shaded region marks the bounds for the age estimate. For most of the initial separation range given by the x-axis, a $Q$ of $3\times10^5$ to $2\times10^6$ is inferred for the planet, with the best estimate being $10^6$, though up to $3\times10^6$ is reasonable if the moon initially started closer to its current location.}
\label{tmig}
\epsscale{1.0}
\end{figure}

We integrate Equations (\ref{spin}), (\ref{seps}), (\ref{sepm}) with the torques given by Equations (\ref{CPLtorques}) and (\ref{CPLtorquem}) to obtain the moon separation $a_m$ as a function of time for Kepler-1708 b-i. Inverting this gives the migration time as a function of $a_m$. We repeat the integration for various values of $Q$, ranging from $5\times10^4$ to $3\times10^6$ and plot the results in Figure~\ref{tmig}. Each black curve represents the migration time to the exomoon's current best estimated location, $a_m=5.54\,a_\mathrm{R}$ for a constant $Q$ displayed above the curve. The shaded magenta region represents the age of the system, $3.16\pm2.26$\,Gyr, where the best estimate is a darker dashed line. For an initial planet-moon separation of about $2\,a_\mathrm{R}$, the $Q$ of the planet can be constrained to $3\times10^5$ for the lower age bound of $\sim1$\,Gyr or about $2\times10^6$ for the upper age bound of $\sim5$\,Gyr. The best estimate for this initial separation is $Q=10^6$, consistent with the analytic result obtained above.

The migration timescale curves change little up to an initial separation of $\sim4\,a_\mathrm{R}$. Thus, whether or not the moon was formed close to the planet near $2\,a_\mathrm{R}$ or further out up to twice that value, the above results still hold. The formation history of the moon will be discussed in more detail in Section~\ref{formation}.

For these calculations we chose an initial spin period of 6 hours for the planet. Because the CPL model only cares about whether $n_m-\Omega_p<0$ and not the exact value of the difference, the actual value of the spin period does not affect the calculation. It is reasonable to assume that the planet will always spin faster than the moon is orbiting during the time period we are considering (just as Jupiter and Saturn in our own solar system; \citealp{mardling}; \citealp{batygin18}). At the minimum separation we consider, $a_m=2\,a_\mathrm{R}$, the moon orbits every 24 hours so it is unlikely that the planet will slow down sufficiently for the moon's orbital rate to exceed the planet spin rate. In fact, at a spin rate of 6 hours the planet's spin energy is nearly ten times the moon's orbital energy at $2\,a_\mathrm{R}$ so there is no appreciable change in planet spin throughout the integration.

\subsection{Kepler-1625}
\begin{figure}
\epsscale{1.0}
\plotone{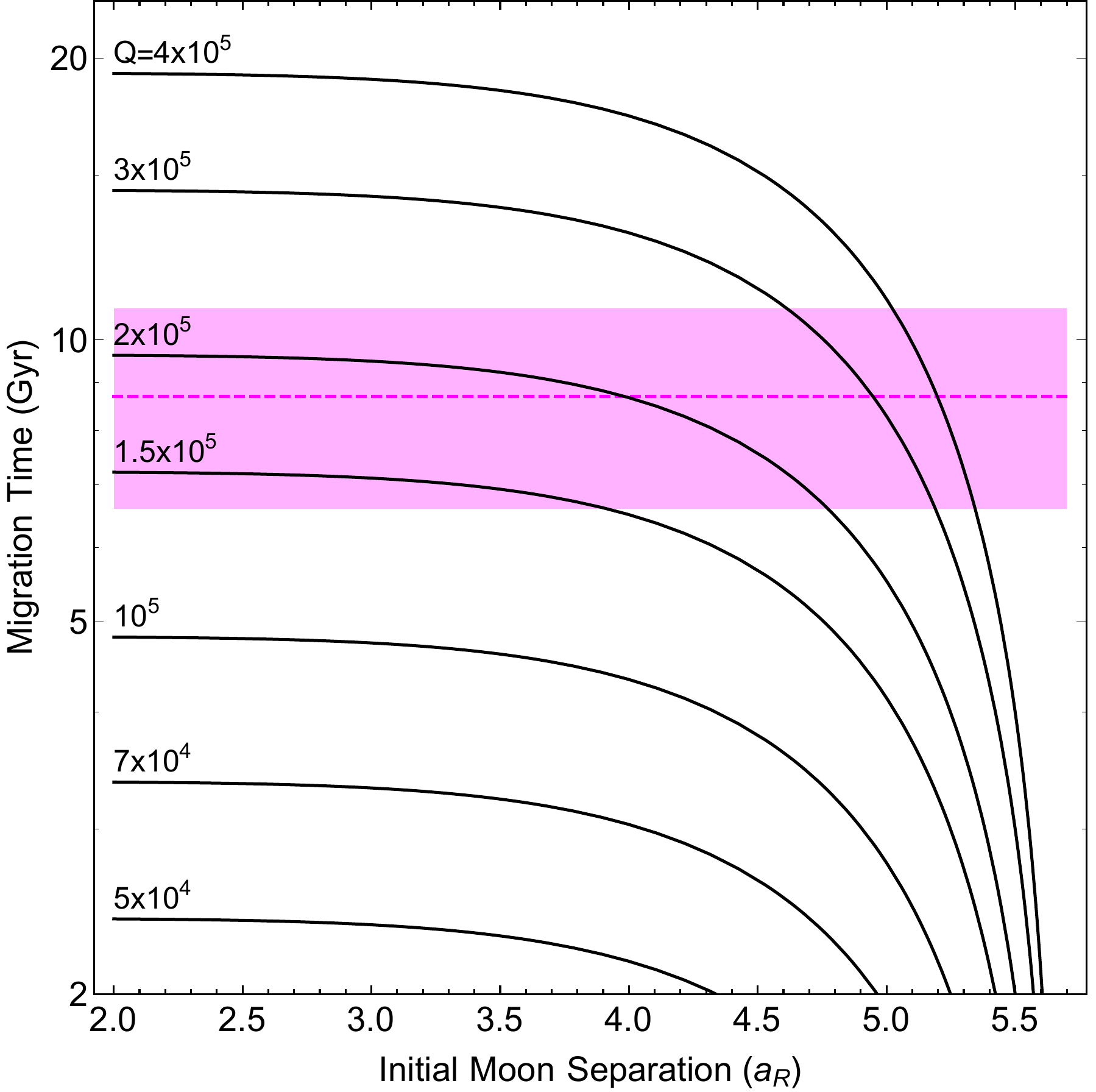}
\caption{Similar to Figure~\ref{tmig}, but for exomoon candidate Kepler-1625 b-i and current separation $a_m=5.7\,a_\mathrm{R}$. The age of the system is about 8.7\,Gyr. For small initial separations, a $Q$ range of $1.5\times10^5$ to $2\times10^5$ is estimated for the planet, but up to $4\times10^5$ is inferred for a moon that initially formed or was captured closer to the current separation.}
\label{tmig1625}
\epsscale{1.0}
\end{figure}

We apply a similar analysis to the exomoon candidate Kepler-1625 b-i \citep{teachey}. As before, we integrate the system numerically using the CPL model to obtain the migration timescales to the current best location for varying $Q$ (Figure~\ref{tmig1625}). In this case, we estimate $Q$ to be between $1.5\times10^5$ and $2\times10^5$, which is lower by a few factors than for Kepler-1708~b. The lower $Q$ may be attributed in part to the wider separation between planet and moon in the Kepler-1625 system. In addition, this system is much older, $8.7\pm2.1$\,Gyr, where the age is reflected by the magenta dashed line and shading. Nevertheless, this range of $Q$ values is still reasonable for Jupiter-sized planets like Kepler-1625~b. If the moon formed closer to its current location and did not migrate an appreciable amount, a $Q$ of $4\times10^5$ or a couple times larger is acceptable, approaching the best estimated $Q$ for Kepler-1708~b. 

We note that some studies place the current moon location much further out, up to $12\,a_\mathrm{R}$ \citep{teachey}. When we consider this possibility, we find that it implies $Q\approx2000$, which is much smaller than what is typically attributed to such planets. This means that for a wide moon orbit, the moon would have had to have formed nearly right at its current location or that such wide orbit solutions are inconsistent with what we know about tidal dissipation. In fact, given the high inclination of the moon's orbit, it is unlikely the moon formed in a circumplanetary disk and may have been captured on a wider orbit instead, close to its present separation \citep{teachey20}.

\subsection{Moon Formation Scenarios}
\label{formation}

The three main formation pathways for moons are in-situ disk accretion as in the case of the Galilean moons of Jupiter \citep{makalkin}, direct impact by a large body like for the Earth-Moon system \citep{cameron}, and satellite capture which may explain the origin of Mars's small moons. These scenarios may each lead to a different starting point for the exomoon candidate. Thus by tracing back its orbital evolution history, we can obtain estimates for $Q$ and the structure of the planet given the moon formation model.

\textit{Disk Formation}. In the in-situ disk model of moon formation, a proto-moon originates in the circumplanetary disk  which it accretes from as it grows \citep{szulagyi}. The study of such moons created by this method generally involve relatively small moons like those around Jupiter, but if the protoplanetary disk of the original planet is large enough, the circumplanetary disk will be massive and can potentially form giant moons, as is demonstrated for the exomoon candidate around Kepler-1625 b \citep{moraes}. In this case, the moon is estimated to have originated between 20 and $50\,R_p$, which corresponds to 3.5-8\,$a_\mathrm{R}$. This is consistent with the simulations performed for the Galilean moons, with the resulting satellites ending up around 4-12\,$a_\mathrm{R}$ from Jupiter, although a few form as a close as $2.5\,a_\mathrm{R}$ \citep{batygin20}. For Kepler-1708~b-i, these estimates place the moon within the ranges proposed in Figure~\ref{tmig}. A $Q$ of $5\times10^5$ to $2\times10^6$ is reasonable for these initial moon separations given the age of the system. For Kepler-1625~b-i, the disk formation model implies $Q=1.5\times10^5-2.5\times10^5$ for the current semimajor axis estimate of $5.7\,a_\mathrm{R}$, but is inconsistent with the larger estimate of $12\,a_\mathrm{R}$. 

\textit{Giant Impact}. Like the origin of our Moon, natural satellites can form following an impact of the planet with another massive body. After the collision, the moon will likely form near the surface of the planet after which tidal forces will push the moon to a more stable orbit \citep{goldreich66}. This results in moon to planet mass ratio estimated to be relatively large, between $10^{-2}$ and $10^{-3}$, for systems of this origin \citep{heller18}. The masses of Kepler-1708~b and its exomoon are not constrained but reasonable estimates such as assumed in this paper will place the moon to planet mass ratio at about $2\times10^{-2}$, roughly consistent with the impact origin scenario. Since this results in moon formation just exterior to $a_\mathrm{R}$, the $Q$ estimate for this theory ranges from $3\times10^5$ to $10^6$. For Kepler-1625~b-i the giant impact scenario gives a smaller $Q$ range of $1.3\times10^5-2\times10^5$.

\textit{Capture}. A binary interaction between two planets or proto-planets can lead to the capture of a secondary body around the planet. The satellite could be captured as close as a single Roche radius but the estimated separations for the capture scenario are not constrained as well as for in-situ disk formation. Generally, capture results in a highly eccentric orbit after which tides in the satellite will circularize the orbit before the outward migration stage \citep{hamers}. The eccentricity damping timescale is less than 100\,Myr for Kepler-1708~b-i and less than 200\,Myr for Kepler-1625~b-i, which is an order of magnitude shorter than the migration timescales considered here so that the moon would have already begun its recession. A similar formation scenario is pull-down capture, which occurs while the planet is still accreting so that the planet and moon grow together as the moon is captured into an orbit around the planet \citep{hansen}. Future more detailed simulations of how Kepler-1708~b or Kepler-1625~b could have captured their moons could be helpful for better constraining the initial values of $a_m$ and in turn the associated $Q$ values of the planets.
 
\section{Conclusion}
\label{sec:conclusion}
We have shown that the tidal evolution of an exomoon can give clues about the host exoplanet's interior structure and tidal dissipation rate. We used recent exomoon candidates Kepler-1708~b-i and Kepler-1625~b-i to calculate theoretical migration timescales of the moon to its current location. The key result is Equation~(\ref{QQ}), which provides a planet's $Q$ for given system parameters. Then we numerically integrated the star-planet-moon system using the CPL model with a resulting $Q$ of $10^5-10^6$ which categorizes the planets Kepler-1708~b and Kepler-1625~b as gas giant planets similar in composition to Jupiter. Although some studies place the moon of Kepler-1625~b at a larger semimajor axis of $\sim12\,a_\mathrm{R}$, we find this would result in an unusually low $Q\sim2000$ unless the moon began at essentially this same distance.

We then considered moon formation scenarios and analyzed the cases of disk formation, giant impact, and satellite capture. We found that if the moon formed in the disk or by impact, the resulting initial planet-moon separation would then evolve to its current value in timescales consistent with $Q$'s ranging from $1.5\times10^5$ to $2\times10^6$. This agrees with our results from numerical calculations that consider a range of starting positions for the moon. We do not rule out the capture scenario but do not gain information about the planet structure in this formulism since initial moon semimajor axis is not well constrained in this case.

As more exomoons are inevitably detected in the future, the methods described here can shed light on the host exoplanet's composition, a property that is generally challenging to constrain. Furthermore, the study of how these moons formed will allow for a better estimate of initial orbital conditions, important parameters for accurately calculating $Q$ and other physical properties. Indeed, in this study we have considered the best estimated parameters, but given the uncertainties in these values, especially for the planet-moon separation and planet and moon masses, better measurements of these will give better constraints on planet interior structure.

The exoplanet radius gap, or Fulton gap, falls in the overlap of planet classes known as sub-Neptunes and super-Earths \citep{fulton2017}. An exomoon detection around a planet that falls in this range of radii could help distinguish between a rocky Earth-like host ($Q\sim10$) and gaseous Neptune-like one ($Q\sim10^3)$. In this way, the hunt for exomoons could be a crucial factor in understanding exoplanetary structure.

\acknowledgments
We thank Alex Teachey for helpful feedback on a previous version of the manuscript. We acknowledge that support for this work was provided by the 2021 Carnegie Institution for Science Venture Grant program. AT acknowledges support from the USC-Carnegie fellowship. 

\bibliographystyle{yahapj}

\end{document}